# The Galaxy Distribution

*homogeneous, fractal, or neither?*

Joseph L. McCauley[+]
Department of Physics
University of Oslo
Box 1048, Blindern
N-0314 Oslo

and

Institute for Energy Technology
Box 40
N-2007 Kjeller

## Abstract

From the standpoint of theoretical physics we can treat Newtonian cosmology as a problem in nonlinear dynamics. The attempt to average the density, in search of a method of making contact between theory and observation, is replaced by the more systematic idea of coarsegraining. I also explain in this context why two previous attempts at the construction of hierarchical models of the universe are not useful for data analysis. The main ideas behind two older competing data analyses purporting to show evidence from galaxy statistics for either a homogeneous and isotropic universe in one case, and for a mono-fractal universe in the other, are presented and discussed. I also present the method and results of a newer data analysis that shows that visible matter provides no evidence that would allow us to claim that the cosmological principle holds, or that the universe is fractal (or multifractal).

In other words, observational data provides us with no evidence that that the universe is either homogeneous and isotropic, or hierarchical.

+ permanent address:    Physics Department
                        University of Houston
                         Houston, Texas 77204 USA

## 1. The cosmological principle

Modern cosmology begins with Einstein's locally-Lorentz invariant geometric theory of gravity and Hubble's law. Cosmology may be described imprecisely as the study of the gross dynamical behavior of the matter in the universe. To give this statement meaning one must define what can be meant by "gross behavior". Cosmologists usually introduce the idea of averaging, where the size of the volume averaged over is left unspecified. I shall use the idea of coarsegraining instead, which arises naturally in modern nonlinear dynamics. We can understand the cosmological principle as the expectation that the universe should be homogeneous and isotropic at a "large enough scale of averaging or coarsegraining".

Another way to express the content of the cosmological principle is to assume that the matter density is spatially constant at fixed times. This idea requires that we define an averaging, or coarsegraining, of the underlying dynamics because the homogeneity assumption, if true at all, can only be true at the most coarse level of global description of the dynamics. The reason for this restriction follows from figure 1, which suggests that the galaxy distribution is not *locally* homogeneous and isotropic.

The cosmologic principle appears superficially to resemble a principle of relativity, but the cosmological principle is not the basis for the general theory of relativity (which is not a "general" relativity principle either) nor is it required by any other known law of physics. The cosmological principle is not an independent law of nature. Many cosmologists and other scientists believe in the cosmological principle, but I shall explain why the assumption of a homogeneous and isotropic universe is not supported by the analysis of the available data on galaxy distributions. This point is not new [1]. What *is* new is that the observational data also provide evidence against a mono-fractal distribution of the galaxies, but provide no evidence for a multifractal distribution.

**2. Coarsegraining and stability in Newtonian cosmology**

The problem of defining average solutions in theoretical cosmology, which is a branch of nonlinear dynamics, is unsolved in general relativity although some progress has been made in Newtonian cosmology. I begin with the idea of coarsegraining a Newtonian cosmology. Coarsegraining is developed systematically in modern nonlinear dynamics [2,3]. This approach is adequate for discussing data analysis in search of either fractal or nonfractal distributions of matter, including multifractal or uniform distributions.

According to Newtonian dynamics we can consider a pressureless dust obeying the coupled quasilinear equations of hydrodynamics [4]

$$\frac{d\vec{v}}{dt} = \frac{\partial \vec{v}}{\partial t} + \vec{v} \cdot \nabla \vec{v} = -\frac{1}{\rho} \nabla \Phi$$

$$\frac{\partial \rho}{\partial t} + \nabla \cdot \rho \vec{v} = 0$$

$$\nabla^2 \Phi = 4\pi\rho \qquad . \quad (1)$$

We can also discuss the galaxy motions as an N-body problem in Newtonian mechanics, where each galaxy is approximated by a point particle (a speck of dust). This picture requires that the intergalactic distances are all large compared with the spatial extent of each galaxy. In any attempt at a treatment of the dynamics where the precision of the level of description is left unspecified and is potentially infinite, the density would be given by the pointwise expression

$$\rho(\vec{x},t) = \frac{1}{N} \sum_{i=1}^{N} \delta(\vec{x} - \vec{x}_i) \quad (2)$$

where $x_i$ is the position of the $i^{th}$ galaxy in the N-body problem. Instead of (2) and N-body dynamics we want to use average or coarsegrained densities in hydrodynamics. We explain below why hydrodynamics is in this case mathematically-equivalent to a certain coarsegrained dust particle dynamics.

A systematic approach to a hierarchy of increasing levels of precision in dynamics and hydrodynamics is made possible by introducing a hierarchy of coarsegrained densities, as is done in modern nonlinear dynamics. At the crudest level of description we divide space, at a fixed time t, into $N_n \ll N$ cubes of size l on each side (the language used here is admittedly Newtonian). Consider next the empirically-observed distribution of galaxies: the number of galaxies found in cell i is denoted by $n_i$. The density in that cell is then $\rho_i = n_i/l^3$, and the average density, at any finite level n of precision of description, is given by

$$\langle \rho \rangle = \frac{1}{N_n} \sum_{i=1}^{N_n} \rho_i \chi(\vec{x} - \vec{x}_i) \quad (3)$$

where $\chi_i$ is a set function that equals unity on the set where the density is $\rho_i$ and is otherwise zero. The idea, if we could carry it out mathematically, would be to solve equations (1) with the coarsegrained density (3) as initial condition and then study the stability of the solution. Clearly, this can be carried out in principle for any degree of coarsegraining. To date, this has only been done for the uniform distribution [4], where $n_i$ is the same for all cells.

It is claimed without proof in [1] that fractal distributions cannot be treated by classical mathematical methods of analysis, that the renormalization group is necessary. This claim is false. The formulation of coarsegraining described above applies to both fractal and nonfractal distributions of dust particles, and the renormalization group approximation is unnecessary (there are no critical point singularities in the coarsegrained picture). More to the point, while the renormalization group can be understood as providing a systematic method of coarsegraining near a critical point, that method has not been successfully extended to cover nonlinear dynamics far from thermal equilibrium, or away from transitions to chaos.

We don't need a continuum interpretation of the dynamics described by (1). We can instead discuss the dust via an appropriate particle dynamics at any coarsegrained level of description of the density. The reason for this is that the method for analysing and solving the quasilinear partial differential equations (1) is via solving for their characteristic curves [5,6], which are generated locally by the differential equations

$$\frac{dt}{1} = \frac{dx_k}{v_k} = \frac{dv_k}{-\frac{\partial \Phi}{\partial x_k}}. \quad (4)$$

These are simply the equations of Newtonian mechanics: each dust particle obeys Newton's law in the gravitational field defined by the other N-1 dust particles. We can rewrite the characteristic equations in the form

$$\dot{x}_k = v_k$$
$$\dot{v}_k = -\frac{\partial \Phi}{\partial x_k}, \qquad (4b)$$

or as Newton's second law

$$\ddot{x}_k + \frac{\partial \Phi}{\partial x_k} = 0 \qquad (5)$$

in 3-space. We can think of streamlines in the 6-dimensional (x,v) phase space (phase flow picture) if and only if the solutions x and v are finite for all real finite times (in which case all singularities of power series solutions of (5) are confined to the complex time plane). That is, the streamline picture holds iff. (5) has no spontaneous singularities, which is not the case in Newtonian cosmology [7,8]. Here, ideas based on caustics and jet-space [9] have provided a useful approach to the nonlinear dynamics.

The cosmological principle would hold if two conditions would be satisfied empirically. First, we would need find a large enough length scale l where the number $n_i$ of particles per cell is roughly the same. This would require that there are no voids and no clusters at that scale of observation. Second, the resulting uniform distribution would have to be effectively stable over a time scale that is not too small on the order of the age of the universe. In part 3 I discuss whether any evidence for a uniform density at large scales l is provided by the observed distribution of the galaxies.

Newtonian cosmologies require a torsion-free flat space (a flat space is one where Cartesian coordinates exist globally because the curvature vanishes everywhere). A flat space can be realized in two ways: (i) as an infinite unbounded Euclidean space, or (ii) as a finite unbounded space in the form of a Euclidean 3-torus. In the former case uniform distributions in Newtonian cosmologies are known to be unstable [4]. In the second case (which is equivalent to solving (1) with periodic boundary conditions) the uniform distribution is stable under certain conditions [4]. In other words, perturbation theory tells us that the cosmological principle can't hold in an infinite Newtonian universe. The extension of Heckmann's analysis beyond the results in his papers shows that the cosmological principle cannot be satisfied even to zero$^{th}$ order in perturbation theory in an infinite Newtonian universe [8].

The Hubble expansion is inferred nonuniquely from redshift data. The galaxy distribution is obtained from the analysis of data derived from redshift data. In the usual Hubble expansion interpretation of redshifts the inferred matter distribution of the observable part of the universe is implicitly presumed to be uniform [7,8], where galaxies accordingly are assumed to recede from one another with a radial velocity field

$$v = H(t)r \quad . \qquad (6)$$

Here, H(t) is the very slowly-varying Hubble 'constant' and is given approximately by H(t)≈100 h km/s.Mpc, with h≈.5 to .6 in our present epoch. Since H(t) varies with time the presumed equivalent observers are accelerated relative to one another. If the universe is not globally homogeneous and isotropic then equation (6) cannot be correct in detail.

I turn next to the observed distribution of galaxies and ask whether there is any evidence to support either the cosmological principle or a hierarchical universe, e.g, a fractal universe. As background for the references for the next section, where observational data are discussed, it may be useful to know that one parsec is 3.2615 light years and that the average distance from the earth to the sun is denoted by one AU. With p as parallax in seconds of degrees, r as distance in parsecs, and L = 1 AU then p ≈ tanp = L/r = 1/r where the units of r are AU per second of degrees.

### 3. Old and new analyses of the galaxy distribution data

I begin by reviewing the standard analysis of the observed distribution of the galaxies, which is described in Peebles' textbook [10]. There, it is assumed without question that the cosmological principle must hold. Next, I review Pietronero's criticism of that analysis, which is essentially correct. Pietronero goes on to argue to the contrary that the universe is a simple mono-fractal. The controversy over the question whether the universe homogeneous and isotropic or fractal continued for the last ten years. I explain below why Pietronero's analysis of the galaxy distribution is also in error, and disagree in the end with both camps: I will explain why there is no evidence *at all* from the observed galaxy distribution that permits the conclusion, at this time, that the universe is either homogeneous and isotropic, or fractal.

**3a. The Standard Model analysis**

"... it may be that the universe is inhomogeneous on all scales, like the velocity distribution in a turbulent fluid: we might have galaxies, clusters of galaxies, clusters of clusters of galaxies, and so on. This is the hierarchical model ... Optical astronomers have searched for inhomogeneity in the distribution of galaxies, and their results indicate some superclustering on scales of 30-50 Mpc. It is possible with radio astronomy to reach out to much larger distances, and on these larger scales it does appear that the universe is at last homogeneous. There is no compelling reason to consider the hierarchical model any further. Indeed, there is good reason not to: we do not know how to incorporate it into a theoretical framework within which we can interpret observational data."

       M. Berry, in <u>Principles of cosmology and gravitation</u> [11]

From this standpoint the cosmological principle is treated as necessity, and other viewpoints (like hierarchical models) must first prove themselves (somehow) in order to be entertained as serious possibilities. The reason for this viewpoint is based partly on philosophy and partly on convenience: the cosmological principle follows from the simplest possible solution of Einstein's field equations [12,13] (the solution is completely integrable and can be found analytically as well), and forms the basis for the analysis of observational data as well as the underpinning for the Standard Model of the universe [10,13]. Without this exceedingly simple solution there is no *global* theory of cosmology. Worse, if the universe would be hierarchical then there is no known analysis of the redshift data that is self-consistent: we do not yet know how to generalize (6) correctly to allow for hierarchical and other universes. Summarizing, if the cosmological principle could be shown to be false, then cosmology would not the coherent body of knowledge that many theorists believe that it is.

The cosmological principle can be obtained theoretically from general relativity by insisting that the metric g and the mass-energy tensor T are globally invariant in a maximally-symmetric space [12,13]. The practical consequence is that the resulting model universe is isotropic and homogeneous, so that the density is spatially constant. This density, as we explained in part 1, must be assumed to be a coarsegrained quantity at some yet unknown "largest scale l" of coarsegraining. Can we deduce such a scale l from the known galaxy distribution data? Is there a scale l of coarsegraining where the observed voids and clusters (see figure 1) disappear, so that the distribution looks homogeneous and isotropic? Is there a crossover from (the observed) voids and clustering to homogeneity and isotropy? As Pietronero has emphasized, figure 1 encourages one to say "No!", so what is the basis in data analysis for the argument to the contrary?

Instead of coarsegraining we consider the following idea from statistical mechanics. Starting with the pointwise density (2) of the dust particles, imagine performing an unspecified average over an unspecified scale l and consider the density correlation function [10]

$$G(\vec{r}) = \langle \rho(\vec{r}_i)\rho(\vec{r}_i + \vec{r}) \rangle. \quad (7)$$

Assume next that we can write the correlation function in the form

$$G(\vec{r}) = \langle \rho \rangle^2 \xi(r) + \langle \rho \rangle^2. \quad (8)$$

This is the form of the density correlation function for a liquid in thermodynamic equilibrium [14] if $\xi(r) \approx 0$ for $r \gg r_o$, where $r_o$ is the correlation length and $\langle \rho \rangle$ is the average density of the liquid, which is macroscopically homogeneous and isotropic. The usual liquid correlation function, with short-distance oscillations reflecting approximately the short-range order of a crystal, does not describe turbulence or any macroscopic or microscopic chaotic behavior besides thermodynamic equilibrium. There is no convincing evidence to suggest that galaxies interacting gravitationally are approximated by the highly unlikely condition of thermal equilibrium. The liquid correlation function is attractive to adherents to the Standard Model because, applied to our dust particles, it allows a crossover to a statistical distribution that obeys the cosmological principle when $r \gg r_o$. According to Davis and Peebles [10,15] the galaxy data for 1 Mpc/h ≤ r ≤ 10 Mpc/h yield $\xi(r) \approx Ar^{-\gamma}$ with $\gamma \approx 1.7$ to 1.8. In other words, there is scale invariance at short separations of galaxies. Fluids in thermal equilibrium away from criticality do not show scale invariance at short distances, while fluids at criticality (where scale invariance *does* hold) have $r_o = \infty$. The correlation length $r_o$ is arbitrarily defined [10] by the nonstatistical mechanical condition $\xi(r_o) \approx 1$, yielding $r_o \approx 5$ Mpc/h for galaxies. Pietronero [1] pointed out the obvious criticism: if you merely look at the data visually (figure 1) then it is clear that both voids and clusters are much larger than the deduced "correlation length" of 5 Mpc/h.

Clustering and voids with sizes that are much larger than 5 Mpc/h tell us that the correlation length, if there is one at all, is much greater than 5 Mpc/h. Where is the error in the standard analysis?

**3b. The argument for a simple fractal model of clustering**

"Although there are wide divergences of view as to the significance, the necessity, and the logical position of [the cosmological principle], the agreement as to its validity is very remarkable, and its utility is beyond doubt."

<div style="text-align: right;">H. Bondi, in <u>Cosmology</u> [16]</div>

Consider instead what is directly measurable by box counting, the number n(r) of galaxies within a sphere of radius r,

$$n(r) \approx \int G(r) d^3 r \quad . \qquad (9)$$

G(r) is the correlation function and n(r) is called the correlation integral. However, do not assume that G(r) becomes constant at $r \gg r_o$. Just count the number of galaxies in the range [0,r], which is actually what the analysis of the last section was really based on anyway. The early result, according to the Pietronero school of thought [1], is universal scaling out to the sample size R of the observational data,

$$n(r) = cr^{3-\gamma} = cr^\nu \qquad (10)$$

with $\gamma \approx 1.2$ to 1.3. This result of earlier analyses [1] is supposed to be true for galaxies in the range 1 Mpc/h $\leq r \leq$ 150 Mpc/h. The authors later reported (the equivalent of) $\gamma \approx .9$ to 1.3 [17] for the CfA1 catalog with .8 to 20 Mpc/h, and $\gamma \approx .8$ for the 1.2Jy catalog, also over only one decade in log-log plots. As they explain [1], the reason for the error in the standard analysis is that $r_o$ is not a correlation length. Instead, $A = r_o^{-\gamma}$ is the amplitude of the correlation function: their analysis suggests that $r_o$ is on the order of magnitude of R, the size of the sample [1].

In contrast with the usual expectation [10,11], the galaxy distribution seems to show scale invariance out to the size of the available samples, with no evidence at all of a crossover to homogeneity and isotropy. Furthermore one can give the result a geometric interpretation.

We can write the correlation integral as

$$n(r) = \frac{1}{N} \sum_{i=1}^{N} n_i(r), \qquad (11)$$

where

$$n_i(r) = \frac{1}{N} \sum_{i \neq j}^{N} \theta(r - |\vec{r}_i - \vec{r}_j|) \qquad (12)$$

is the number of galaxies within a sphere of radius r that is centered on the ith galaxy. This is a more useful form for data analysis and follows from (60) for a discrete distribution. If we would find that $n(r) \approx r^\nu$ then, roughly speaking, ν is the correlation dimension [18] $D_2$. It is easy to prove that $D_2 < D_H$ where $D_H$ is the Hausdorff dimension [3,18] of the support [19] of the galaxy distribution. According to the Pietronero school the correlation dimension is around 2 and the galaxy distribution is mono-fractal [17]. If it were true then this would provide quantitative underpinning for an old speculation made originally by Mandelbrot [20]. Hierarchical models were also put forth in the prefractal era [21,22].

Some Standard Model anaylses are subject to the following criticism [1]: so-called "buffer zones" are added to the data in order to build in homogeneity (which believers in the Standard Model expect) because the observational data are not homogeneous. "Buffer zones" are not scientific and certainly should not be used. Defenders of Standard-Model-thinking retort that the Pietronero analysis uses unfair sampling [23]. Both schools of thought are right in their criticism of each other.

The fractal model enthusiasts [1,17] have *not* included in their analysis the broken circles in the hypothetical galaxy sample shown as figure 2. They counted only the points inside circles that are completely within the data cone (excepting an illegal plot showing three decades of scaling where they violate their own advice [19] by using the equivalent of *only* broken circles). Advocates of the Standard Model correctly point out that this makes the galaxy sampling unfair [23]: in plots of log n(r) vs. log r, the points near the center of the cone dominate the large r part of the plot, weighting the galaxies near the center of the sample unfairly against those near the edges. In other words, one might just as well plot the log-log graph for $n_i(r)$ for the central point and forget the rest of the data, so far as the fractal analysis of ref. 1 is concerned. This criticism must be taken seriously.

**3c. The correlation integrand**

" ... if a sample contains too few points there may be no way to get any information about it. In such a case one has to wait for better (observational) data."
P.H. Coleman and L. Pietronero, in <u>The Fractal Structure of the Universe</u> [1]

According to the most recent report by the proponents of a fractal universe [17] the correlation integral is supposed to scale like $n(r) \approx cr^\nu$ with $\nu \approx 1.7$ to 2.1 from .8 to 20 Mpc/h for volume-limited samples in the CfA1 catalog (the older result [1] was $\nu \approx 1.5\text{-}1.7$). A homogeneous and isotropic universe would obey $n(r) \approx dr^3$, whereas two-dimensional sheets would require $n(r) \approx br^2$. The Standard Model advocates claim scale invariance with $\nu \approx 1.2$ to $1.3$ for galaxies in the range $1$ Mpc/h $\leq r \leq 10$ Mpc/h with a crossover to homogeneity at large scales. A model of the universe that is scale invariant at short distances, but where the cosmological principle holds "at large enough scales" [24] is described very simply by $n(r) = cr^\nu + dr^3$. According to the analysis of the last section there is, as yet, no observational evidence for a finite value of the prefactor $d$. I will also explain below why there is no evidence for a universal value of the scaling index $\nu$, or for nonuniversal scaling either.

Scale invariance means simply that $n(\lambda r) = \lambda^\nu n(r)$. The solution of this equation is $n(r) = cr^\nu$. The function $n(r) = cr^\nu + dr^3$ is not scale invariant unless $v = 3$. The sum of any number of exponentials with different exponents is not scale invariant. However, data that are not scale invariant can easily give the appearance of scale invariance if the data analysis does not extend over enough decades on a log-log plot. An example is shown as figure 3. A rule of thumb that is generally accepted in critical phenomena (the famous "Geilo Criterion") is that one needs to demonstrate scaling over at least three decades in a log-log plot. This criterion is hard to satisfy and is necessary, if not sufficient. Let us ask next: what are the possible ways that a function can be approximately scale invariant?

A very simple possibility is that $n_i(r) = c_i r^v + \delta n_i(r)$ where $\Sigma \delta n_i(r) = 0$. As an example the fluctuations $\delta n_i(r)$ may be given by Poisson noise. An example of Poisson noise is to take $\delta n_i(r) = d_i r^3$ where $\Sigma d_i = 0$. Poisson noise is assumed, but not demonstrated, in the analysis described in part 3b above. I will explain below why the available data do not agree with this assumption.

Another possibility is that local scale invariance holds,

$$n_i(r) = c_i r^{v_i}, \quad (13)$$

and there is a single term that dominates so that

$$n(r) = \frac{1}{N} \sum_{i=1}^{N} n_i(r) \approx r^{v_i} \quad (14)$$

yielding $v = v_i$. Here, the other terms do not cancel each other (as in Poisson noise), they are merely small when compared with the dominant term. This would also be consistent with the analysis of 3b, and we will ask whether the data agree with this assumption.

Still another possibility is that local scaling (64) holds, so that the $i^{th}$ galaxy has a correlation summand $n_i(r)$ that is scale invariant, but scale invariance does not hold globally. This would be the case if, for example, the spread in local exponents $\nu_i = (\ln n_i(r))/\ln r$ obtained from the data is large. In fact, all three cases listed here can be analysed simply by studying the correlation "integrand" (summand) [19], by plotting $\ln n_i(r)$ vs. $\ln r$ for as many galaxies as is possible in a sample.

In order to provide a more careful analysis of the question of scale invariance Martin Kerscher has studied the correlation integrand $n_i(r)$ under the following restriction (see [19]): we do not compute the correlation integrand for galaxies centered on the broken circles in figure 2, only for the solid circles, but (in contrast with [1,17]) when computing $n_i(r)$ we weight all included galaxies $i = 1,2,3,..,N_{max}$ exactly the same. This means that we use the same maximum allowed radius $r_{max}$ for every galaxy for which $n_i(r)$ is computed, so that $r_{max}$ is limited to about 10 Mpc/h ($N_{max} \ll N$). This limits the possible local scaling indices $\nu_i$ to galaxies that are not close to the edge of the sample, weights them all the same, but admittedly is still not completely free of the criticism of unfair sampling. However, this analysis represents the best that can be done with the current limited data. *Even with this restriction* we find that the *spread* in local exponents $\nu_i$ determined by the best fit of a scaling law $\nu_i \approx \ln(n_i(r))/\ln r - \ln c_i$ to the local slopes is too large [19] (figures 4 and 5) to allow us to characterize the sample by a scaling index $\nu$. Figure 4 represents a only small selection, fifteen of the ninety-two plots based on the CfA1 catalog [19], showing that the local slopes peak near $\nu \approx$ 1.7-1.8 (the 1.2 Jy catalog is also analyzed in [19]). The deviations from the maximum value, for all ninety-two plots [19], is summarized by figure 5. Even if one were inclined to ignore the considerable spread in local slopes, the log-log plots are only over one decade, not enough to evidence for scale invariance (see figure3).

The available galaxy data are not adequate to claim that global scaling exists, or even that local scaling exists with variable exponents $\nu_i$: there are simply not enough decades available on the required log-log plots. To test reliably for scaling one would need data out to at least 1000 Mpc/h. The present data may be consistent with local scaling with a broad spectrum of local exponents $\nu_i$, but certainly not with global scale invariance of the correlation integral with a single exponent $\nu$. The data might be multifractal but, again, testing this assumption would require data out to at least 1000 Mpc/h. Note that there is no evidence at all for either homogeneity and isotropy (requiring $\nu \approx 3$), or for a crossover to homogeneity and isotropy. In other words, there is no evidence from visible matter to support the cosmological principle. Instead, one sees voids and clustering (but not scaling) out to the present limit of reliable data.

The problem with the plots arising from data analyses of references [10, 15] and [1,17] can be stated in another way: correct error bars were never shown. Had correct error bars been exhibited it would have been seen that there was no scientific basis for a controversy between the two camps. This point was made earlier by the author [25].

## 4. Newtonian dust dynamics in hierarchical cosmologies

Discussions of the cosmological principle can be laid to rest for the time being (and perhaps forever) because the idea cannot be tested. Global cosmological predictions cannot be tested, given our limited observational knowledge. A more reasonable alternative to is to try to build models that reflect the available data represented by figure 1, which must be regarded as *local*. Monofractal scaling is out. Multifractal scaling cannot yet be tested, but it is still of interest to build hierarchical and more general types of coarsegrained models that agree with the dust distribution illustrated in figure 1. Many different multifractal and nonmultifractal models will be consistent with the (inadequate) data corresponding to figure 1. This is not different from the present situation in turbulence [26,27].

Wertz [22] proposed a non self-consistent Newtonian model of a hierarchical universe. His global hierarchy consists of spherical clusters of different sizes. To zeroth order the clusters are treated as noninteracting, and within each cluster the dust particles obey Hubble's law with different Hubble constants. Each cluster is therefore assumed to be an isolated Heckmann model, to zeroth order. Within a cluster the dust particles are distributed both homogeneously and isotropically. This approach disagrees strongly with observation (see figure 1), where the data show no evidence for homogeneity locally. The hierarchical aspect of Wertz's model is confined to clusters of clusters. There, Wertz's prediction cannot be tested globally due to grossly inadequate observational data. Therefore, a completely different model is needed, one that attempts to model the local data of figure 1. In such a model the global prediction for the density is unimportant because it cannot be tested anyway.

An approach that is free of the criticism of Wertz's model was provided by Ribeiro's swiss cheese model [28,28,30] in general relativity. The swiss cheese model is supposed to be locally hierarchical but is globally homogeneous and isotropic at large distances from the (artificial) singularity, corresponding in spirit to $n(r) = cr^v + dr^3$, which is pleasing to believers in the Standard Model. Ribeiro's model is based the density formula proposed by da Vaucouleurs' [31], "$\rho(r)$" $\approx r^{D-3}$ with D<3, which introduces an artificial (because nonphysical) singularity into the analysis at r = 0 but is otherwise everywhere smoothly differentiable. This artificial singularity motivated the (also incorrect) claim [1] that hierarchical distributions cannot be treated by classical (i.e., nonlinear) dynamics methods, and require instead (an unknown, because never defined) renormalization group method of analysis. Ribeiro treats the singular local density via a Tolman model. Assuming that the cosmological principle holds, he then matches the result as inner solution to a Friedmann dust model as outer solution.

The worst aspect of this model is that the inner Tolman solution based on the smooth (except at r=0) da Vaucouleurs density "$\rho(r)$" $\approx r^{D-3}$ does not at all represent the local structure shown in figure 1, where voids have sizes on the order of the sizes of clusters. By writing $n(r) = cr^D$ and assuming that "$\rho(r)$" $= n(r)/4\pi r^3$ we would certainly obtain nothing other than "$\rho(r)$" $\approx r^{D-3}$, but this formula cannot be used to represent a hierarchy because it implicitly and artificially smooths the big fluctuations that are the voids. *da Vaucouleurs' formula merely replaces the real matter distribution, which is* *erratic with large fluctuations due to voids, by a smooth one that has the same correlation integral n(r).* The only way to do the dynamics correctly, if we pay attention to figure 1, is to use (instead of "ρ(r)") a correct coarsegraining of the dust.

We know that the distribution of galaxies within a cluster is not homogeneous, and also is not mono-fractal. The available observational data are inadequate to test hierarchical models globally because present information on clusters of clusters is grossly inadequate to test for scaling. Therefore, a sensible aim for the present might be to try to construct a local model of a hierarchical universe. Here I mean local in the spirit of Heckmann, where boundary conditions on the gravitational potential (the global aspect) are ignored, but (in contrast with Heckmann and Wertz) where the clustering of dust particles shows clustering and voids within the 'island universe' (cluster). It is necessary to depart completely from Wertz, therefore, by treating as possibly-hierarchical what he treated as isotropic and homogeneous, and then completely to ignore the global aspect (clustering of clusters) that he treated as hierarchical. Such a treatment could incorporate the voids and clusters shown in figure 1, which neither Wertz's nor Ribeiro's model does.

Invariance of the local velocity field [8] cannot be used as the basis for a local hierarchical model, because that assumption leads to a smoothly varying local density rather than to coarsegrained densities arranged hierarchically on a tree [2,3,26]. At this stage no one knows how to produce an analytic example of a Newtonian model that is locally hierarchical or more general. For hierarchical (piecewise constant) matter density, the hydrodynamics equations likely must be studied numerically for stability.

## Acknowledgement

This article is based in part lectures given at both the Institute for energiteknikk and the University of Oslo in the spring of 1997 during my sabbatical, during which time I enjoyed guestfriendship provided by both institutions. I am very grateful to my hosts, Jan Frøyland and Arne Skjeltorp, and also to Finn Ravndal and Øyvind Grøn, for many pleasant and stimulating discussions. Earlier in my sabbatical, from September 1996, through February 1997, I was guest professor in Lehrstuhl Wagner at Ludwig Maximillian's Universität in München, where I contributed to the research reflected in part 3c. I am grateful to Herbert Wagner, Thomas Buchert, and Martin Kersher, without whose help I could not have entered into the interesting and stimulating debate on galaxy clustering. The figures were provided by Martin Kerscher, who also performed the data analysis. This work was supported financially by The Research Council of Norway, the Institute for Energy Technology (at Kjeller, Norway), and the Departments of Physics at the Universities of Oslo (UIO), München (LMU) and Houston (UH).

**Figure Captions**

1. A portion (summation of slices) of the largest three-dimensional survey of the distribution of galaxies.

2. A sketch of a hypothetical galaxy sample, illustrating which galaxies are used in ref. 6 (broken circles are included) and which are used in ref. 14 (solid circles only). Adding 'buffer zones' amounts to manufacturing points that do not belong to the data within the part of the broken circle that lies outside the sample.

3. Log-log plot of the function $f(r) = cr^{-\gamma} + dr^{-\eta}$ (stars) with $\gamma = 1$ and $\eta = 0$ together with the function $f(r) = 18r^{-\gamma'}$ (dashed line) with $\gamma' = 0.9$.

4. Fifteen (of ninety-two [19]) plots of $\ln n_i(r)$ against $\ln r$ for the volume limited CfA1 sample with 40 Mpc/h depth (solid line) are shown. The dotted line is the best fit to a local power law $n_i(r) = c_i r^{\nu_i}$. The long dashed line would represent $n_i(r) \approx r^2$, and the short dashed line represents $n_i(r) \approx r^3$. The data are too sparse to permit more than one decade on a log-log plot.

5. The frequency of local exponents $\nu_i$ for the data represented by all ninety-two plots of local slopes in [19] for the volume-limited CfA1 catalog. The distribution of local slopes peaks near $\nu \approx 1.7$-$1.8$ and shows considerable spread.